\documentclass[twocolumn,floats,aps]{revtex4}
\usepackage{times,amsmath,graphics,psfrag,latexsym}

\begin{document}
\title{Stable Vortex Configurations in a Cylinder}
\author{R.J. Zieve and L.A.K. Donev}
\affiliation{Physics Department, University of California at Davis}
\begin{abstract}
We calculate stable arrangements for a single superfluid vortex pinned to the wall
of a stationary cylindrical container.  We find that, independent of the details
of the pinning site, stable vortices must subtend most of the cell
horizontally and cannot be vertical or nearly vertical.  More generally,
the geometry of a container severely limits the possible vortex configurations,
making macroscopic trapped vortices less common than previously believed.
\end{abstract}
\maketitle

Vortex lines in superfluid helium have been studied for over forty years 
\cite{hell}, but many fundamental questions remain.  Moving vortices feel a
drag force, known as mutual friction, from interaction with the superfluid
excitations.  The effects of mutual friction have been thoroughly measured in
systems of many vortices \cite{fricreview}, but there is still no microscopic
theory of the interaction. Similarly, the interaction of two closely approaching
vortex lines is unknown.  Another unresolved issue is the structure of the
vortex core, and whether it can change size and shape to alter the energy and
momentum of a vortex.  Other questions involve superfluid turbulence, which 
can be described
\begin{figure}[b]
\begin{center}
\scalebox{.25}{\includegraphics{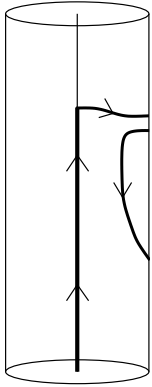}}
\caption{\small Vortex partially trapped around a wire.  The picture
illustrates the possible interaction of the vortex with a second vortex
pinned to the wall of the chamber.}
\label{f:vortwire}
\end{center}
\end{figure}
as a tangle of vortex lines.  In a decaying tangle, the mean vorticity has the
same power-law dependence on time as for a classical system, even at such low
temperatures that the normal fluid density is negligible \cite{Smith,
McClintock}. In addition, experiments \cite{Tabeling} and simulations 
\cite{LTscaletheory} show that superfluid  turbulence follows a Kolmogorov
scaling law at low temperatures.   The power law, prefactor, and even the
deviations from ideal scaling are the same for superfluid and classical
turbulence.  How the quantized vortices in the superfluid mimic so well the
behavior of classical viscous vortices is unknown, as is the length scale at
which differences appear.  A better understanding of single vortex behavior may
help answer these questions.  

A few years ago, an unusual method for observing vortex motion was discovered
\cite{3helicop, 4helicop}. A vortex line is partially trapped around a thin
wire, continuing through the fluid as a free vortex, as shown in Figure
\ref{f:vortwire}.  Although the vorticity is initially created by rotating the
cryostat, the circulation around the wire is long-lived, and measurements can
be made after the cryostat is brought to rest.  In the case of a vortex
partially attached to the wire, the self-velocity drives the free end around
the cell.  Its motion is detected through changes in the normal modes of the
wire's vibration, which can identify the position where
the free end attaches to the wire to within 1\% of the wire length.  The 
system also lends itself to computer
simulation, with detailed computations of the vortex motion showing excellent
agreement with the experimental data \cite{Schwarzheli}.  The  vibrating wire
technique allows direct investigation of how a single vortex interacts with
various situations, including other vortices in the container, heat pulses,
geometrical obstructions, or fluid flow across the cell. One of  the simplest
interactions is with a single additional vortex line, pinned to the cell
wall in a metastable configuration. The work discussed here began as an attempt
to identify metastable vortex configurations for use in
the experiments.  Since the
measurements, unlike most vortex studies, will be made with the cryostat
stationary, we use this situation in our calculations.  From here on, we refer
to metastable pinned vortices in a stationary cylinder as ``stable."  We find
that the only stable single vortex configurations are mainly horizontal, an
orientation that may be difficult to achieve after rotation around a vertical
axis. This result may be part of the reason for the smooth vortex motion
observed in the experiments: the vortex line is not perturbed by interactions 
with macroscopic pinned vortices because few if any such vortices are present.

\section{Vortex configurations}
We test for stability with a computer program along the lines of Schwarz's code
for modeling vortex dynamics \cite{Schwarz}.  The equations used are discussed 
at length in  Ref. \cite{Schwarz}, and in subsequent work by other groups
\cite{simSamuels, simTsubota, simAarts, Aartsthesis, simqucl}, so they are 
summarized only briefly here.  We treat the vortex cores as massless and thin; 
that is, the core diameter is small compared to the curvature along the core. 
The equation governing the motion of superfluid vortices is essentially the
Euler equation for incompressible fluid flow, with an additional term for
mutual friction between the vortices and normal fluid.  Physically, in the
absence of friction, the vortex cores move at the local superfluid velocity. 
In our code, a set of points along a vortex core is stepped through time, with
the velocity ${\bf \dot{s}}$ of a  point on the core given by 
${\bf \dot{s}}={\bf
\dot{s}_o}+\alpha({\bf \dot{s}_o}\times{\bf s^\prime})$, where
\begin{align*}
{\bf \dot{s}_o}=\ &\frac{\kappa}{4\pi} \int^\prime \frac{({\bf s_1}-{\bf s})
\times d{\bf s_1}}{|{\bf s_1} -{\bf s}|^3} +
\frac{\kappa}{4\pi}\ln\frac{2\sqrt{l_+l_-}}{e^{1/4}a_o} 
({\bf s^\prime}\times{\bf s^{\prime\prime}})\\ &+{\bf v_a}+{\bf v_b}.
\end{align*}  
Here $\alpha$ is a friction coefficient, $\kappa$ is the quantum of
circulation, and $a_o$ is the radius of the vortex core. In ${}^4$He, $\kappa =
9.98\times 10^{-4} \mbox{cm}^2/\mbox{s}$ and $a_o = 1.3\times 10^{-8}$ cm. The
first term integrates over the vortex core to find the contribution to 
velocity of the non-local parts of the vortex.   We approximate the core as
linear between each pair of specified points. When calculating this term at a
given point, we omit the vortex segments on either side of that point.  These
segments, of lengths $l_-$ and $l_+$, instead contribute to the second term,
which arises from the local curvature, $R$.  
${\bf s^\prime}$ is the unit vector
along the direction of the vortex core, and ${\bf s^{\prime\prime}}$ is a
vector of length $1/R$,  directed toward the center of curvature.  The applied
velocity is ${\bf v_a}$; in our work this comes from either a vertical vortex 
located  at the center of the cylinder, or a constant vertical velocity.  The
boundary field ${\bf v_b}$ satisfies $\nabla^2{\bf v_b}=0$ everywhere inside
the cylinder, and must cancel the perpendicular velocity at the cell wall from
the other terms to ensure that the entire velocity field has no perpendicular
component at the wall.   

Before presenting our program and results, we briefly discuss vortex
pinning.  A standard picture is that surface roughness
can pin a vortex by providing a local minimum in the vortex's length,
and hence in its energy.  Schwarz gave this intuition numerical support
by showing that a hemispherical bump on an infinite plate will trap
one end of a half-infinite vortex line \cite{Schwarzbump}.  His work includes 
no explicit
pinning term in the equations of motion; the important point is that the
geometry itself can catch the vortex line.  As the vortex moves nearby,
the distortion of the fluid flow by the bump directs the vortex towards
the bump.  The vortex typically spirals around the bump before reaching
a stationary configuration.  The particular geometry used 
allows an exact solution to Laplace's equation using image vortices.
Furthermore, two bumps on opposing parallel walls pin vortices \cite{Schwarz}.
That is, a vortex with its endpoints on the two bumps will reach a stable
position even if there is an imposed superfluid flow between the plates.
At some critical velocity, the vortex comes off the bumps and is swept
along by the flow.
Without the pinning sites, the vortex would always move with the flow.

\begin{figure}[b]
\begin{center}
\scalebox{.3}{\includegraphics{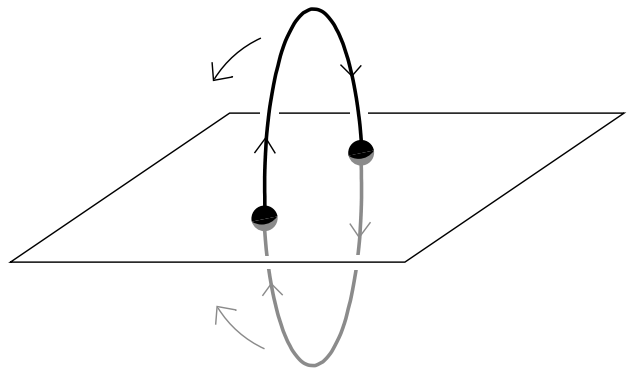}}
\caption{\small Vortex with both ends pinned to bumps on an infinite plane.
The vortex (black) and its image (gray) attract each other and ultimately 
annihilate.}
\label{f:planebumps}
\end{center}
\end{figure}

As an example of pinning sites that do not support a stable vortex, consider
two bumps on an infinite plane, as shown in Figure \ref{f:planebumps}.  A
vortex arching in a semicircle between the bumps, with its image vortex, forms
a ring.  Its own velocity pushes it along its axis, but with the ends pinned
the vortex and image collapse towards each other.  They then attract each other
more strongly, and the vortex eventually hits the wall and annihilates.   

More generally, a vortex pinned in any container will be attracted to the
nearest wall.  The velocity contribution of the vortex itself may cancel
the influence of the wall, resulting in a stable vortex configuration,
but whether this happens depends on the container shape and the locations
of the vortex endpoints.  Here we find the possible stable vortices
terminating on the curved walls of a cylinder. Except when otherwise
stated, the calculations use a cell of radius 0.01 cm and length 0.5 cm.

\section{Numerical considerations}
We use a fifth-order Runge-Kutta-Felberg method with a variable time step for
solving the ordinary differential equation.  The step size is determined
by the relative
difference between the fourth-order and fifth-order steps, except that when the
velocity is very small the absolute difference is used instead.   This greatly
improves stability and eliminates the need for the hopscotch algorithm Schwarz
cites \cite{Schwarz}.  With our code, a vortex ring of radius 0.01 cm
propagates for hundreds of centimeters without noticeable distortion, even in
the absence of friction.  The time step is about $2.5\times 10^{-3}$ seconds
for 25 points along the vortex core. The time step is limited by stiffness and 
is roughly quadratic in the minimum point spacing, so using a stiffly stable
method (e.g., backward Euler) could allow a larger time step.  By contrast,
without friction Schwarz could propagate a vortex this size only 0.7 cm before
an instability appeared with a time step of $2\times 10^{-4}$ seconds, and this
distance decreases with longer time steps \cite{Schwarz}.  A similar improvement
in performance was noted with a Crank-Nicolson algorithm and variable time step
\cite{Aartsthesis}.  As a test case for stability of various algorithms,
Aarts \cite{Aartsthesis} considers a distortion with
period two points on either a vortex ring or a straight vortex.  Even with no
mutual friction, these distortions do not grow with our algorithm.  

\begin{figure}[bht]
\begin{center}
\psfrag{xcoord}{\scalebox{2}{\small $x$ coordinate of vortex center ($\mu$m)}}
\psfrag{time}{\scalebox{2}{\small Time (seconds)}}
\psfrag{angle}{\scalebox{2}{\small 0.7$\pi$}}
\scalebox{.5}{\includegraphics{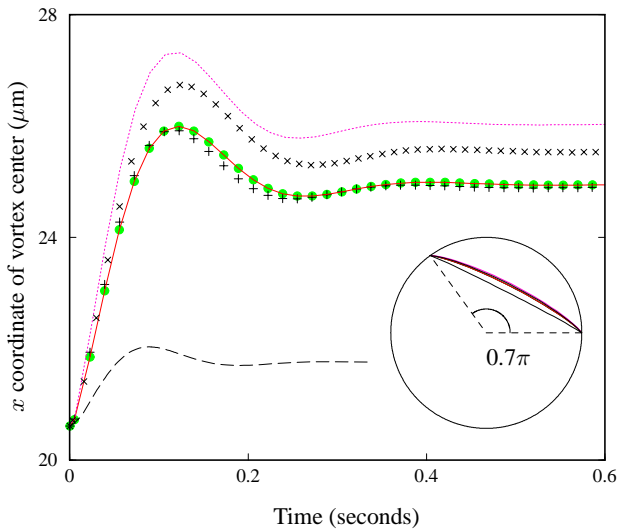}}
\caption{\small Motion of the middle of a vortex line,  calculated
using different boundary treatments.  All use $\alpha=0.5$ and cell
radius $0.01$ cm, and all but the solid line have cell height $0.5$ cm.
The symbols show inversion images followed by Fourier transforms on a
$16\times 16$ (x), $32\times 32$ ($\bullet$), or $64\times 64$
(+) grid.   The lines show inversion images only (dotted),
infinite radial continuations of vortices (dashed), and inversion images
with Fourier transforms on a $64 \times 64$ grid for a cylinder 1 cm long
(solid).  The inset shows the corresponding final configurations.}
\label{f:fftcomp}
\end{center}
\end{figure}

A minor difference in the calculation is that Schwarz approximates vortex
segments by the circle passing through three consecutive points.  We do use the
radius of this circle as the local curvature of the vortex line. However,
rather than calculating the tangent to the circle, we use the vector between a
point's nearest neighbors as the direction of the vortex at that point.  We
note that even using the exact tangent to the inscribed circle provides a
consistent picture of the vortex line only locally, since the local
calculations for two consecutive points assume different shapes for the
intermediate segment.  Since the tangent vector is used only to define the
direction of the friction contribution to the vortex motion, our approximation
is identical to that of Schwarz for a stationary configuration.  The
two are nearly equivalent even for vortex dynamics except in situations of
uneven point spacing or sharp bends.  
We did verify with a direct calculation that our numerics are essentially
independent of this calculational detail.

\begin{figure}[htb]
\begin{center}
\psfrag{th}{\scalebox{2}{$\theta$}}
\psfrag{psi}{\scalebox{2}{$\psi$}}
\scalebox{.4}{\includegraphics{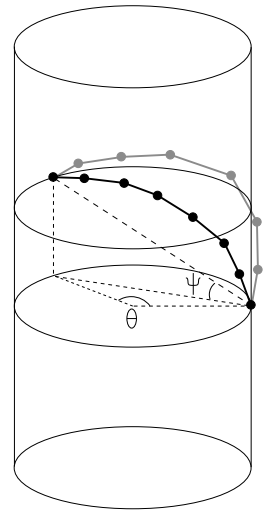}}
\caption{\small Construction of image vortices for cylinder by inversion.
The vortex (black) ends on the cell wall.  The nodes
are separately inverted, and their images define the core of the image
vortex (gray).  The angle $\theta$ is the horizontal separation of the
vortex endpoints, measured from
the cylinder axis.  The angle from the horizontal of the line connecting
the endpoints is $\psi$.}
\label{f:inversion}
\end{center}
\end{figure}

We make one departure from the differential equations by subtracting the
velocity tangent to the vortex core before updating the configuration.
This prevents points along the vortex line from drifting towards one
end of the line.  We believe this is reasonable even for dynamical
simulations, and it definitely should not affect the equilibrium
configurations we discuss here.  
Since our vortices are nearly stationary, the point spacing remains
constant to within a few percent once we neglect the tangential
velocity component.  Our code checks the point spacings at each time
step, and can add or delete points to keep the spacings uniform.  No
points were ever added or deleted during the computations presented here.

For many purposes, the boundary contribution ${\bf v_b}$ to the velocity
field can be ignored.  However, stability depends heavily on the flow field
near the container wall, where ${\bf v_b}$ can be significant. As an
illustration, Figure \ref{f:fftcomp} compares the motion of the center of a
vortex line for several different treatments of the boundary. The pin sites lie
in a horizontal plane at the center of the cylinder, and are separated by an
angle $\theta=0.7\pi$.   Straight radial extensions of the vortices are the
furthest from giving a vanishing perpendicular velocity at the walls, and
result in substantially different motion from the other treatments.  Assigning
images by inversion is a better approximation.  The original vortex is
described by discrete points along its core.  As shown in Figure
\ref{f:inversion}, we invert each of  these points in a horizontal circle with
the same radius $R_c$ as the cylinder.  Thus, in cylindrical coordinates,
$(\rho,\theta,z)$ maps to  $(R_c^2/\rho,\theta,z)$.  These image points are
then used to describe the core of a new vortex line.  In this way any vortex
with both ends terminating on the cell wall becomes, with its image, a closed
loop. Since the image vortices give only an approximate solution, we then use a
Fourier series solution of Laplace's equation to cancel any remaining
perpendicular velocity at the curved wall.  We calculate the perpendicular 
velocity arising from the real and image vortices at a grid of points on the
curved surface of the cylinder.  We do a two-dimensional discrete  Fourier
transform of these velocities to get the coefficients of a Fourier series 
solution of the boundary value problem.  Subtracting this solution from the
other velocity contributions helps meet the boundary condition.  An infinite
Fourier series would give the unique solution to the fluid velocity for the
given vortex core arrangement, but the discrete Fourier transform we use
can be only an approximation.
The
perpendicular velocity is cancelled exactly at the grid points and
approximately between points.   Figure \ref{f:fftcomp} shows results from
several grid spacings.  As the Fourier grid becomes finer, the combination of
inversion images and Fourier solutions converges. We do not reduce the grid
size beyond the point spacing along the vortex core.  We assume there is no
flow through the top and bottom faces of the cylinder, a reasonable
approximation as long as the vortices are far from the ends.  The solid curve
of Figure \ref{f:fftcomp}, which represents calculations in a longer cylinder,
verifies that we are in the regime where vortex motion does not depend on cell
length, and justifies our neglect of the ends.  The $xy$ projection of the
final  vortex configurations for the various treatments are shown in the inset
of Figure \ref{f:fftcomp}.  Except for the straight vortex continuations, the
final arrangements cannot be resolved on the scale of the graph.

\begin{figure}[b]
\begin{center}
\includegraphics{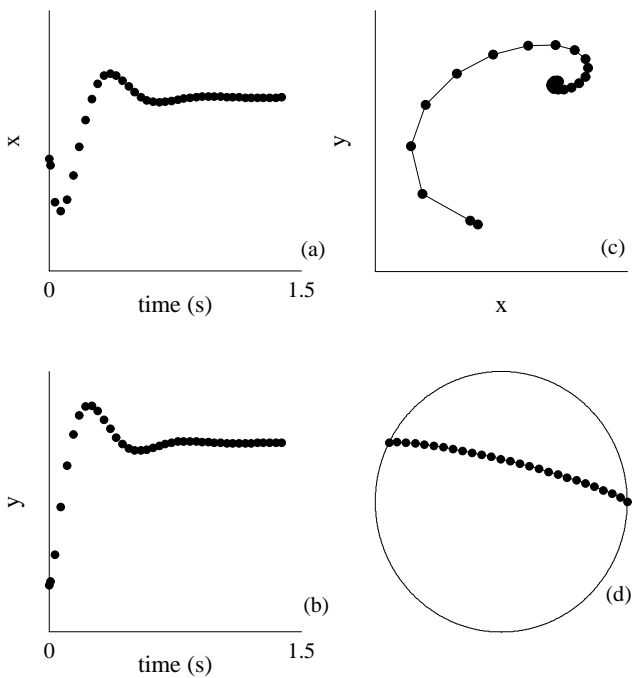}
\caption{\small Vortex line settling into a stable configuration.  The $x$ and $y$
coordinates of the middle of the vortex line both show damped oscillations 
as a function of time in ($a$) and ($b$).  In ($c$), the projection 
of the middle of the vortex onto the $z=0$ 
plane, parametrized by time, spirals towards its equilibrium position.
Finally, ($d$) shows the projection onto
the $z=0$ plane of the cylinder and the final vortex configuration.
The pin sites are at an angle of $0.85\pi$ in the $z=0$ plane and at
an angle of $0.25\pi$ from the horizontal, and the friction coefficient is
$\alpha=0.5$.}
\label{f:stablevort}
\end{center}
\end{figure}
\begin{figure}[t]
\begin{center}
\psfrag{y (microns)}{\LARGE $y$ coordinate of vortex center ($\mu$)}
\scalebox{.5}{\includegraphics{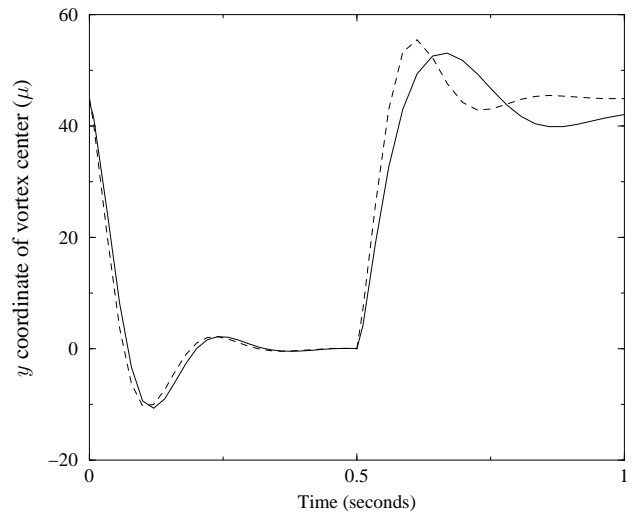}}
\caption{\small Decaying oscillations on a vortex stretched between two planes.
The plane separation is 0.02 cm, the friction coefficient is 0.5, and there
are 25 evenly spaced points along the vortex.  The applied velocity is zero
up to 0.5 seconds and -0.7 cm/s in the $x$ direction thereafter.  The dashed
lines represent the analytic solution for waves on a straight vortex.}
\label{f:Kelvinosc}
\end{center}
\end{figure}

To study pinning, one might introduce bumps on the wall of a cylinder and work
with the resulting boundary field.  With this approach, the point spacing along
the vortex would have to be small compared to the bump size, and the boundary
solution would be complicated by the bumps themselves.  Instead, we impose 
pinning externally,
requiring that the endpoints of the vortex remain fixed at certain points on
the cylinder wall.  We then use a relatively large point spacing.  As long as
the size of the pin site is small compared to the point spacing, the lack of
detail near the site should not affect the vortex behavior.  After describing
typical stable and unstable vortex behavior, we will
present additional justification of this procedure.

When the chosen endpoints support a stable configuration, the vortex line
spirals towards it.  Figure \ref{f:stablevort} shows the behavior  of a vortex
as it moves toward a stable arrangement, and the final vortex configuration. 
Tracking the location of a given point on the core as a function of time shows
decaying oscillations in each coordinate.  These oscillations are 
the lowest frequency Kelvin waves.  For an infinite straight vortex,
the mode with wavelength $k$ has $\omega=\kappa k^2 A/4\pi$, where $A \approx
\ln(\frac{1}{ka})$, and time constant $\tau =1/\alpha\omega$
\cite{Kelwave}.  In Figure \ref{f:stablevort}, the oscillations have
$\omega=10.76$ rad/s and decrease in amplitude by a factor of 0.0436
per cycle.  Since the endpoint separation is 0.026 cm, the lowest mode
has $k=\pi/0.026=120$ cm$^{-1}$, corresponding to $\omega=15.48$ rad/s
and damping of $e^{-\pi}=0.0432$ per cycle.

\begin{figure}[htb]
\begin{center}
\includegraphics{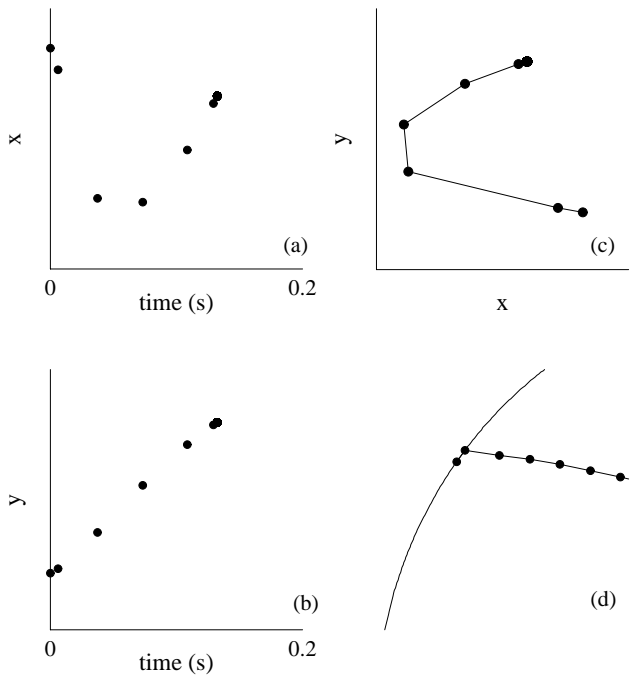}
\caption{\small Vortex line being pulled into wall.  As a function of time, the 
$x$-coordinate of the middle of the vortex ($a$) appears to begin oscillating.
However, the $y$-coordinate ($b$) moves steadily in one direction.  At
a time less than 0.2 seconds, the time step becomes too small to observe
further time evolution.  In ($c$) the $z=0$ projection, parametrized by time,
is shown.  Part ($d$) shows the portion of the final configuration near one of 
the cylinder walls, with the first segment of the vortex pulled against the 
wall.  The pin sites are at an angle of $0.8\pi$ in the $z=0$ plane and at
an angle of $0.27\pi$ from the horizontal, and the friction coefficient is
$\alpha=0.5$.}
\label{f:unstablewall}
\end{center}
\end{figure}

The damping per cycle is in excellent agreement with theory, and 
the discrepancy in the
frequencies arises from boundary effects.  We 
find a similar effect for the parallel plane geometry, as shown in Figure
\ref{f:Kelvinosc}. Up to time 0.5 seconds, there is no applied velocity field. 
The vortex line is initially bent, in its equilibrium configuration for a
constant applied velocity -0.7${\bf\hat{x}}$ cm/s.  The first oscillations
show the return of the vortex to vertical. The $y$-coordinate of the
vortex center is shown.  At time $0.5$ seconds the applied velocity is
changed to -0.7${\bf\hat{x}}$ cm/s, and the subsequent oscillations
show the vortex settling into a bent configuration. The frequency
is significantly faster in the first half second, when the vortex
is oscillating around a straight line.  The dotted lines show the
exponentially damped oscillations expected from the analytic solution
for waves on a straight vortex, with the initial amplitude matched to the
simulations and no other adjustable parameters.  The remaining discrepancy
in the first half second arises because the image vortices needed to meet
the parallel plane boundary conditions result in a scalloped extension
of the vortex, rather than a vortex with a sinusoidal distortion.
For a nearly straight vortex, the two differ only slightly, while for
the bent vortex the difference is more significant.  We have checked
that our simulations for waves on a long free vortex do agree with the
analytic dispersion relation.  Thus we are confident that the observed
decrease in frequency for the pinned vortex in a cylinder is caused by
the boundary and the vortex curvature.

The damped Kelvin oscillations of Figure \ref{f:stablevort} appear only
for certain endpoint positions.
For other locations at least one coordinate
shows no oscillations, and the time step decreases until it reaches the minimum
allowed value.  After this, the progression in time is so slow that the further
development of the vortex cannot be reached.  Figure \ref{f:unstablewall}
illustrates this possibility.  The final point shown in each of the first
two parts of the figure is actually a large number of data points, which cannot
be distinguished because of the small time step.
Examining the calculation at each point shows
that the time step crash arises near where the vortex joins the wall.  The
velocity at this spot becomes large and strongly dependent on the exact
configuration.  Often, in fact, part of the core tries to move through the
cylinder wall.  We interpret the time step crash as indicating that an actual
vortex line would be pulled into the wall by its image vortex until it
eventually annihilated or freed itself from the pin site.  Thus the time step
crash indicates an unstable pin configuration.  If we were using
bumps on the cylinder wall as pin sites instead of externally imposed pinning,
the end of the vortex would move along the bump until reaching the cylindrical
wall, and would then continue along the cylinder.   

We tested our treatment of the pin centers in a configuration of two parallel
planes.  For this geometry an infinite series of image vortices would give an
exact solution to the boundary conditions for the fluid flow.  We sum the
series out to ten times the plane spacing, and from there we continue the
vortices to infinity as straight lines perpendicular to the planes. Schwarz
carried out parallel plane calculations with hemispherical bumps, no external
pinning, and a point spacing small compared to bump size.  He found that the
critical velocity for the vortex to remain on a bump was given by $$v_c =
\frac{\kappa}{2\pi D}\ln\frac{b}{a_o},$$ with $D$ the spacing between planes
and $b$ the bump radius \cite{Schwarz}. Using externally imposed pinning, we
also find that a vortex with one end pinned to each plane reaches a stable
configuration as long as the fluid velocity between the planes is sufficiently
small.  For large fluid velocities, we instead observe the time crash behavior
described earlier, with the vortex attempting to move through the walls.  The
crossover between the two regimes defines a critical velocity, which depends on
the vortex point spacing near the walls but not on the point spacing midway
between the planes.   Our observation of the vortex crashing into the wall is
consistent with earlier work that assumed a vortex becomes parallel to the wall
at the  critical velocity \cite{Hedge}.  A numerical solution of the
Hall-Vinen-Bekarevich-Khalatnikov equations, setting vorticity parallel to the 
wall as a boundary condition, gave good agreement with experimental results on
pinning in rough channels \cite{Hedge}.

Our own approximation has the following interpretation.
For a given applied velocity, any bump with radius larger than some value $b_1$
would pin the vortex.  In addition, for a given point spacing, we can only
expect reasonable results for bumps smaller than some other value $b_2$, which
we expect to be on the order of the point spacing.  For larger bumps, our
neglect of the distortion of the flow field near the bump will lead to incorrect
behavior. If $b_1<b_2$, then our program can do a realistic treatment of a bump
of size $b$ with $b_1<b<b_2$ and find a stable equilibrium position of the
vortex. On the other hand, if $b_1>b_2$, then any bump treated reasonably by
our program would fail to pin the vortex.  Since the program enforces pinning,
the result is an unphysical crash of the vortex into the wall.  To model the
effects of a larger bump, we would have to increase the spacing between our 
points.

\begin{figure}[tb]
\begin{center}
\psfrag{Point spacing}{\LARGE Spacing between points along vortex ($\mu$m)}
\psfrag{Bump radius}{\LARGE Bump radius ($\mu$m)}
\scalebox{.5}{\includegraphics{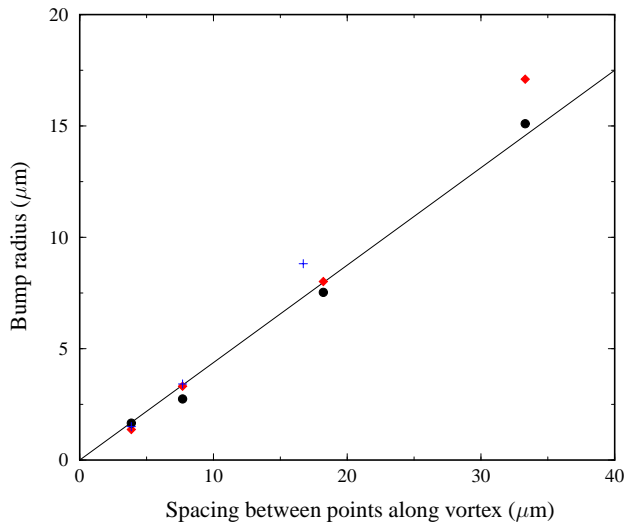}}
\caption{\small Bump radii with the same critical velocity as various point spacings.
The critical velocities were calculated for a vortex pinned between parallel
planes of separation 0.01 cm (+), 0.02 cm ($\diamond$), and 0.04 cm
($\bullet$).
The line is a guide to the eye.}
\label{f:bumpsize}
\end{center}
\end{figure}

To estimate the bump size corresponding to a given point spacing, we use
the parallel plane geometry.  We identify a point spacing with the bump
size which yields the same critical velocity, using Schwarz's formula
to convert bump size to critical velocity.  
We use the initial point spacing along the vortex, but the spacing typically
changes by less than 5\% during a calculation.
Figure \ref{f:bumpsize}
shows this relationship between bump size and point spacing, for three
different plane separations.  The points lie close to a line of slope
0.4, with significant deviations only when the point spacing becomes too
large relative to the plane separation.  As long as the point spacing
is much smaller than the plane separation, the plane separation does
not change the effective bump radius, giving some confidence that
the same correspondence will hold in a cylindrical geometry as well.
Most of the results here correspond to bumps of radius about
3 $\mu$m in our cylinder of radius 100 $\mu$m.  

\begin{figure}[tb]
\begin{center}
\psfrag{Time (seconds)}{\LARGE Time (seconds)}
\psfrag{ycoord}{\LARGE $y$ coordinate of vortex center ($\mu$m)}
\scalebox{.5}{\includegraphics{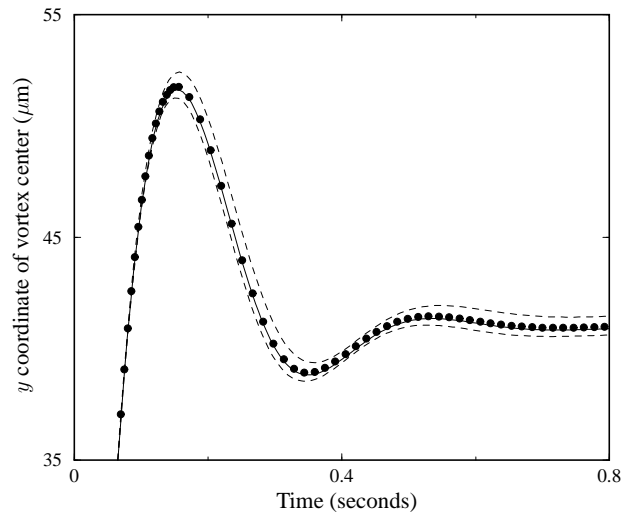}}
\caption{\small Comparison of dynamics for real bumps and for imposed pinning. 
The motion of the middle of a vortex line is shown.  The points show the 
behavior for actual bumps.  The solid line is for imposed pinning, with
point spacing corresponding to the same bump size.  The two dashed lines are
for slightly larger and smaller point spacings.}
\label{f:realbump}
\end{center}
\end{figure}

As an additional check on our treatment of bump size and point spacing,
we have run simulations of a vortex pinned between parallel planes, with
hemispherical bumps as described by Schwarz \cite{Schwarzbump}.  The plane
separation is 0.02 cm, and the bump has radius 7.8 $\mu$m.  We use
point spacing 0.63 $\mu$m near the bump, and the friction
coefficient is 0.5.  The solid circles of Figure \ref{f:realbump} show one
component of the motion of the vortex line's midpoint.  We then repeated
the calculation without the bumps, but with imposed pinning.  The lines
of Figure \ref{f:realbump} represent three different point spacings:
from lowest to highest, 25 $\mu$m, 18 $\mu$m, and 12.5 $\mu$m.  By our
critical velocity matching criterion, the second of these corresponds
to a bump radius of 7.8 $\mu$m; and indeed, the middle, solid trajectory
runs very close to the circles.  The other point spacings yield the 
dashed lines, which are noticeably further from the path of the vortex
on the actual bump.  This demonstrates that our proposed correspondence
between point spacing and bump size gives the correct dynamical behavior
far from the pinning site.

\begin{figure}[b]
\begin{center}
\scalebox{.5}{\includegraphics{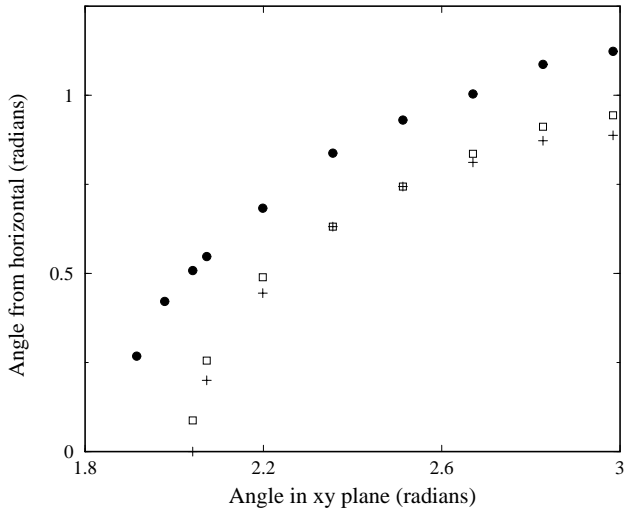}}
\caption{\small Stable regime for single vortex in cylinder of height 0.5 cm and
radius 0.01 cm.  The separation of the two pinned ends is described by the 
angle in the horizontal plane, measured from the cylinder axis, and by the 
angle from the horizontal of a straight line 
between the pin sites.  The three
curves represent a bump size of 3 $\mu$m with friction coefficients 
of 0.5 (+) and 5.0 ($\Box$), and a bump size of 17 $\mu$m with friction 
coefficient 5.0 ($\bullet$).}
\label{f:stablenocirc}
\end{center}
\end{figure}

Our final deviation from the ideal equation is that
in finding equilibrium states, we use an unphysically large
friction coefficient.  Thus the calculated path of the vortex line as it
settles to an equilibrium position does not correspond to real vortex dynamics
in superfluid helium at low temperature.  The final situation, however, should
be unaffected.  The one exception is that a too-low friction coefficient can
lead to a time step crash when a vortex overshoots its equilibrium
configuration while spiraling into position.  In this situation the friction
value is not merely an artifact of the numerics.  Low friction
could also lead to a real vortex line's being unable to reach a stable position.  
As shown in Figure \ref{f:stablenocirc}, this is a small effect, changing 
the boundaries of the stable regime by
only a few percent for a friction coefficient of at least
0.5.  In the data discussed below we use the large-friction limit, since
our concern is to find ultimately stable vortex positions.  In our actual
experiment, with rotation and with many vortices present in the cell, the
conditions for trapping a vortex will be much more complicated than the
present calculations.  For most tests we use a straight vortex as the initial
configuration, but we have also used parabolic vortices, and scaled versions of
stable vortices.  The different initial conditions yield the same ultimate
stable configurations and the same dividing line between stable and unstable.

\section{Results and Discussion}
In the vertical direction, in which a cylinder has no curvature, all
vortices are unstable.  As might
be expected, the vortices suffer the same fate as vortex arches on
a plane.  For horizontal vortices, there is a regime of stability when
the vortex spans much of the cell; for example, a vortex running along a
diameter is stable.  As the angle subtended by the vortex decreases, the
attraction to the cylinder wall grows stronger until the configurations
become unstable.  Introducing any vertical displacement between the
pin sites quickly reduces the stability.  Figure \ref{f:stablenocirc}
shows the stability regime for our typical point spacing of about 7.7
$\mu$m, corresponding to a bump radius of about 3 $\mu$m.  The effects
of changing the friction coefficient or point spacing are also shown.
The larger point spacing is about 33 $\mu$m, corresponding to a 17 $\mu$m
bump radius.  
For larger point spacings, we cannot expect meaningful results because
the spacing is no longer ``small" compared to the cell radius.  For smaller
spacings, an isolated bump of the corresponding size is 
too small to be constructed in the physical experiments.

\begin{figure}[tb]
\begin{center}
\scalebox{.5}{\includegraphics{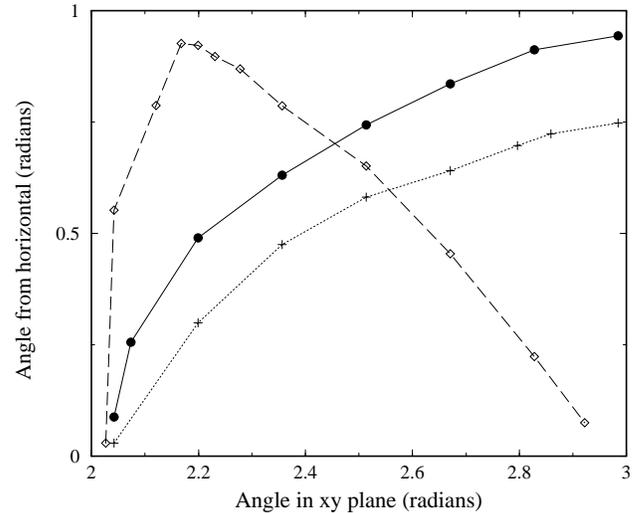}}
\caption{\small Stable regime with and without a central vortex.  The two 
pinned ends are separated both horizontally and
vertically.  The three curves represent no additional fluid flow in the cell
($\bullet$), an  $N=1$ vortex at the center (+), and an $N=-1$ vortex at the
center ($\diamond$).  The vertical component of the pinned vortex is in the
same direction as the $N=1$ central vortex.}
\label{f:stablecirc}
\end{center}
\end{figure}

\begin{figure}[b]
\begin{center}
\scalebox{.7}{\includegraphics{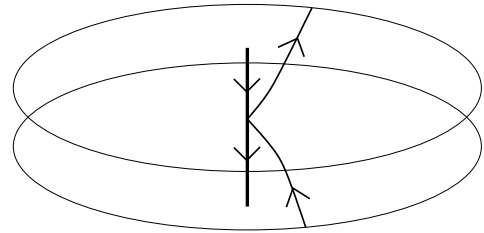}}
\caption{\small Vortex line being pulled in to center of cell.  
The central vortex is directed opposite the vertical component
of the vortex pinned to the outer walls.}
\label{f:unstablecenter}
\end{center}
\end{figure}

In the planned experiments, a single vortex will be trapped around a vertical
wire near the middle of the cylinder.  Thus, in Figure \ref{f:stablecirc} we
examine the stability of the trapped vortex line in the presence of a vortex
along the $z$ axis.  The central vortex has a noticeable effect on stability.  
In the most likely configuration, the central vortex and  the vertical
component of circulation of the pinned vortex have the same direction. Since
the two vortices repel each other, the loop is attracted more strongly
towards the wall and destabilized. However, if the central vortex and 
pinned loop
have opposite circulation, they attract each other.  For horizontal angles near
0.7$\pi$, this pulls the loop away from the wall and helps stabilize it.  As
the pinned vortex traverses more of the cell, though, a portion of it must pass
too near the central vortex.  The $1/r$ velocity field of the central vortex
leads to an instability, attracting the other vortex more and more strongly. 
Again the final result is a time step crash, in this case caused by the large
central velocity field rather than by an attraction to the outer wall.  Figure
\ref{f:unstablecenter} illustrates this  situation, which determines the
boundary of the stable regime for horizontal  angles above 2.2 radians.  

Another issue of experimental importance is having a substantial vertical
component, since typically cryostats rotate only about a vertical axis.
Reproducibly creating a mainly horizontal pinned vortex may be difficult
under these circumstances, so we examined ways to stabilize more nearly
vertical vortices.

Imposing a vertical superfluid flow alters the stability regime, as
discussed in early work on vortices in channels \cite{Glaberson}.   For a
horizontal vortex, the vertical flow changes the curvature and can pull the
vortex away from the wall.  If the vertical velocity is too large, the vortex
hits the wall on the opposite side of the pin site.  For some angles less 
than $0.67\pi$, a
range of vertical velocities can stablilize a vortex.  An angle of 0.6$\pi$,
for example, supports a stable vortex with vertical velocities from 0.03 cm/s
to 0.11 cm/s. However, an external velocity field most strongly affects vortices
perpendicular to it.  Once the vortex moves out of a horizontal plane, the
vertical flow has less stabilizing effect.

We also investigated sharp projections into the cell as potential pinning
sites.  The calculations discussed above ignore the structure of the pinning
center but assume its dimension are small compared to the point spacing.  A
long spike must be treated slightly differently.   We consider two long, thin
spikes, aligned vertically with each other.  The vortex follows the projections
along their lengths, then stretches freely between their ends.  We vary both
the angle between the projections and the wall, and the length of the
projections.  Without circulation at the center of the cell, no parameters
support a stable vortex.  We have tested projections with lengths up to
$\frac23$ the cell radius, at angles from $0.06\pi$ to $\pi/2$ with the wall.
In every case, the center of the vortex line is pulled towards the wall. As it
moves, the portions near the projections bend.  Eventually the vortex forms a
hairpin near the protrusion, and the time step becomes too small to use. 
We believe an actual vortex would be pulled into the wall and destroyed.  A 
central vortex can stabilize the vortex between the
projections, but only when the two have opposite circulation.  Thus long spikes
are probably not useful pin centers for our experimental purposes.

In the actual experiments, a cell is typically 0.15 cm radius and 5
cm long.  As shown earlier, the length is unimportant as long as the
cylinder is much longer than it is wide.  Furthermore, except for the
vortex core size, which appears only in the logarithmic prefactor of the
local term, the governing differential equation scales with distance.
Since the logarithm produces only a small deviation from perfect scaling,
multiplying all length scales by 15 gives reasonable results for a
radius 0.15 cm container.  For perfect scaling, the angles shown in
Figures \ref{f:stablenocirc} and \ref{f:stablecirc} would not change
with cell size.  The maximum bump size represented by the calculations becomes
about 0.25 mm, a practical number for deliberately imposed roughness.

We find that in a cylinder, stable macroscopic pinned vortex lines
have a limited range of possible configurations, all requiring a large
horizontal component.  Similar limits on stable vortex configurations
apply to other geometries as well.  For example, Schwarz noticed that for
parallel planes, pin sites not directly opposite each other can support
a stable vortex, but with a reduced critical velocity \cite{Schwarz}.
As the lateral separation increases, the vortex becomes more nearly
parallel to the wall, a geometry that pulls the vortex towards its image.
We investigated this behavior with our imposed pinning. For a plane
separation of 0.02 cm, we find that vortices with at least 0.04 cm lateral
displacement are unstable even with no applied velocity field.  With real
pin centers, the vortex ends would leave the pins and move closer
together, making the vortex more nearly perpendicular to the planes.

The limited possibilities for stable trapped vortices offer at least a
partial explanation for the extremely smooth vortex motion observed
in experiments \cite{3helicop, helipreprint}.  The cryostat rotates
to create circulation, but stops before measurements begin.  The
additional circulation present in the cell during rotation could pin to
the walls, as shown in Figure \ref{f:vortwire}.  It should then disturb the
motion of the detached portion of the central vortex.  With many
such vortices traversing the cell, the central vortex motion would
be irregular.  Our work here suggests that large pinned vortices are
not common, particularly if the initial circulation is vertical.
The limits on vortex configurations may also be relevant to flow
experiments in channels and to the problem of remnant vorticity
and critical velocities.  Vortices pinned to parallel plates have
been observed in superfluid helium both before and after rotation
\cite{Awschalom}, but other geometries may be less prone to such pinning
of macroscopic vortex lines.  Even if pinned vortices are present,
there may be significant restrictions on their orientation.

Our calculations have no implications for vortices with length comparable to 
the cell roughness, believed to create a vortex mesh covering the wall.   A
vortex pinned between two bumps with radius comparable to their spacing could
behave like a vortex line between parallel planes.  Without
considering the detailed shape of the pinning sites, we cannot address this
length scale.  Instead, the macroscopic vortices we discuss have stability
determined by the large-scale cell geometry.

Since these macroscopic vortices remain far from the wall for most of
their length, their behavior may be similar to truly free vortex lines.
In future experiments, we hope to study their interaction with a
vortex partially trapped on the wire.  We have begun to calculate the
vortex motion in this experimental setup, for direct comparison with
the measurements. In addition, we plan further computational work to
search for stable vortex configurations involving multiple vortex lines
and vortices of the same scale as the wall roughness.  A separate
line of work, of less relevance to our own measurements but more applicable
to other experiments, would be to study the stability of pinned vortices
in a rotating cylinder.

We thank A. Tesdall for his early work on the computer code.

\end{document}